\documentclass[twocolumn,letter]{jpsj2}

\title{%
Pseudogap and Superconductivity in Iron-Based Layered Superconductor
studied by Fluctuation-Exchange Approximation
}

\author{%
Hiroaki \textsc{Ikeda}
\thanks{E-mail:hiroaki@scphys.kyoto-u.ac.jp}
}

\inst{
Department of Physics, Kyoto University, Kyoto 606-8502
}

\recdate{\today}

\abst{%
We investigate interplay between magnetic fluctuations and superconductivity in the effective five-band Hubbard model for iron-oxypnictide superconductors on the basis of the fluctuation-exchange approximation.
As for the normal-state properties, we find the pseudogap behavior in the NMR relaxation rate and the spectral weight in the electron-doped region, while we cannot find such behavior in the hole-doped region.
The pseudogap behavior originates from the band structure effect, that is, existence of high density of states just below the Fermi level.

Solving the superconducting Eliashberg equation, we find that the most probable candidate for the pairing symmetry is the sign-changed $s$-wave spin-singlet state.
For small Hund's coupling $J$, the eigenvalue is not so sensitive to carrier doping, and seems to be irrelevant with antiferromagnetic (AF) spin fluctuation.
We suggest that correlation between spin and spin-quadrupole is important as the pairing mechanism as well as the AF fluctuation.
}

\kword{%
iron-oxypnictide, superconductivity, FLEX, multiband Hubbard model, pseudogap
}

\begin{document}
\maketitle

The recent discovery of the Fe-based layered superconductor LaFeAsO$_{1-x}$F$_x$~\cite{rf:Kamihara} with the transition temperature $T_\mathrm{c}=26$K has promoted highly intensive research activities.
In the new family of FeAs superconductors, $T_\mathrm{c}$ has been elevated to over $50$K by substituting rare earth for La,~\cite{rf:Ren} which is the highest after the high-$T_\mathrm{c}$ cuprates.

Many experimental and theoretical efforts have revealed the following common features.
The undoped system undergoes the stripe-type antiferromagnetic (AF) transition with the structural phase transition.~\cite{rf:Cruz,rf:Nakai}
With electron doping, the system shows the superconducting transition.~\cite{rf:Kamihara}
The hole-doped system~\cite{rf:Wen,rf:Rotter,rf:Sasmal} also have the same order of $T_\mathrm{c}$, which is insensitive to carrier doping and the AF fluctuation.~\cite{rf:Kamihara,rf:Wen,rf:Nakai}
In the band calculation,~\cite{rf:Singh} the dispersion near the Fermi level is mainly composed of Fe $3d$ orbitals, which are occupied by six electrons in the undoped system.
The Fermi surfaces (FSs) are composed of a 3-dimensional one around Z, two hole cylinders around $\Gamma$ and two electron cylinders around $M$.
The dispersion relation and the FSs have been roughly verified by the angle resolved photoemission spectroscopy (ARPES).~\cite{rf:Liu}
As for the superconducting gap symmetry, the NMR Knight shift indicates the spin-singlet state.~\cite{rf:Matano,rf:Grafe,rf:Ning,rf:Terasaki,rf:Kawabata}
$T_\mathrm{c}$ seems to be robust for impurities.~\cite{rf:Kawabata}
The penetration depth~\cite{rf:Hashimoto}, the specific heat~\cite{rf:Mu} and ARPES~\cite{rf:Ding,rf:Kondo} indicate almost isotropic two gap system, while the NMR relaxation rate $1/T_1$ shows no coherence peak and the $T^3$ law,~\cite{rf:Nakai,rf:Matano,rf:Grafe,rf:Ning,rf:Terasaki,rf:Mukuda} which suggest existence of line nodes.
At present, these facts are controversial.~\cite{rf:Nagai}
Conventional theory by the electron-phonon interaction cannot explain high $T_\mathrm{c}$.~\cite{rf:Boeri}
We can expect that unconventional superconductivity originates from the electron correlation.

Among many theoretical research studies, the most promising pairing state is the sign-changed $s$-wave spin-singlet state (hereafter, $s^\pm$-wave state).~\cite{rf:Mazin,rf:Kuroki}
Kuroki {\rm et al.}~\cite{rf:Kuroki} have indicated that the dispersion near the Fermi level can be well explained by simple two-dimensional square lattice of an Fe atom, although two Fe atoms are contained in the actual unit cell.
Using the random phase approximation (RPA), they investigated the effective five-band Hubbard model with the Coulomb interaction on an Fe site;
the intra-orbital Coulomb $U$, the inter-orbital Coulomb $U'$, the Hund's coupling $J$ and the pair-hopping term $J'$.
Recently, by the third-order perturbation theory,~\cite{rf:Nomura} and RPA in 16-band d-p model,~\cite{rf:Yanagi} the possibility of such $s^\pm$-wave state has been verified within weak coupling approach.
In this paper, we investigate interplay between magnetic fluctuations and superconductivity in the above effective five-band Hubbard model on the basis of the strong coupling approach.

First, we obtain a similar band structure by the tight-binding fitting using hopping parameters in {Table~I} of Ref.\ref{rf:Kuroki}.
The band dispersion and the FSs in the unfolded Brillouin zone (BZ) are illustrated in Fig.1.
$\Gamma'$ $(\pi,\pi)$ becomes equivalent to $\Gamma$ $(0,0)$ in the folded original BZ.
Electron pocket around $M$ and hole pocket around $\Gamma$ ($\Gamma'$) reproduce the FSs in the original band structure,
although the FS around $\Gamma'$ is somehow larger than that in the band calculation by Kuroki {\rm et al.}
The $yz$, $zx$ and $x^2-y^2$ orbitals have large weight near the Fermi level.
\begin{figure}[ht]
\begin{center}
\includegraphics[height=33mm]{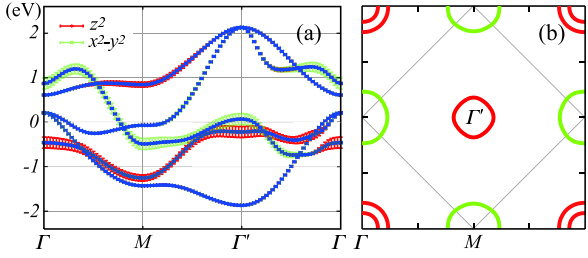}
\end{center}
\caption{The band dispersion (a) and the FS (b) at undoped $n=6.00$ in the unfolded BZ.
In (a), red and green bars represent the weight of $z^2$ and $x^2-y^2$ orbitals, respectively.
In (b), red line around $\Gamma$ ($\Gamma'$) and green line around $M$ represent hole and electron FSs, respectively.
Gray line denotes the original BZ.}
\end{figure}

We here examine the correlation effect for this band structure within the self-consistent second-order perturbation (SC-SOPT), and the fluctuation-exchange approximation (FLEX).
The FLEX calculation in the multiband system follows Ref.\ref{rf:Takimoto}.
In the present five-band model, the normal (anomalous) Green's function $\mathcal{G}_{\ell m}(k)$ ($\mathcal{F}_{\ell m}(k)$) and self-energy $\mathit{\Sigma}_{\ell m}(k)$ can be treated as a $5 \times 5$ matrix in orbital indices, while the spin and the charge (orbital) susceptibilities ($\chi^{s,c}_{\ell\ell',mm'}(q)$) as a $5^2 \times 5^2$ matrix.
With $\hat\chi^0$ simple bubble diagram of $\hat{\mathcal{G}}$, $\hat\chi^{s,c}=\hat\chi^0\pm\hat\chi^0\hat V^{s,c}\hat\chi^{s,c}$, where $\hat V^{s,c}$ denote bare vertices for the spin and charge sectors.~\cite{rf:FLEX}
We set $\hat\chi^s=\hat\chi^c=\hat\chi^0$ in SC-SOPT, and $\hat{\mathit{\Sigma}}=0$ in RPA.
Actual numerical calculations have been carried out mainly with $64 \times 64$ $k$-point meshes and $256$ Matsubara frequencies.
Some calculations have been checked with $64 \times 64 \times 512$.
We hereafter set $U'=U-2J$ and $J'=J$ in the Coulomb interaction, and a unit of energy to be electron volt (eV).

\begin{figure}[ht]
\begin{center}
\includegraphics[height=33mm]{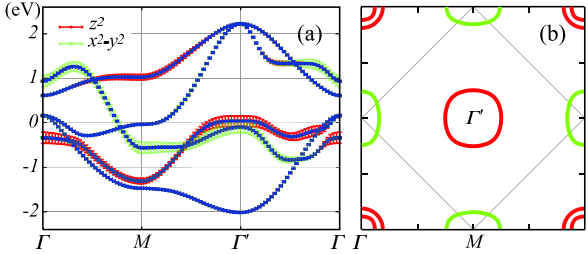}
\end{center}
\caption{The band dispersion (a) and the FS (b) at $n=6.00$ for $(U,J,T)=(1.32,0.22,0.01)$ in SC-SOPT.}
\end{figure}
In Fig.2, we illustrate the band dispersion and the FSs included only the self-energy shift $\mathit{\Sigma}_{\ell m}({\mib k},\mathrm{i}\pi T)$ in SC-SOPT.
Compared with Fig.1, $z^2$ orbital near $\Gamma'$ shifts upward, and $x^2-y^2$ orbital downward.
With large $J$, these shifts increase, and a new FS with $z^2$ orbital appears.
Also in simple SOPT and FLEX, the tendency is the same.
At a glance, this seems to explain a large outer FS observed in ARPES.~\cite{rf:Liu,rf:Ding,rf:Kondo}
In ARPES, however, a flat band, which is assigned to a $z^2$ orbital, is observed below the Fermi level.
In FLEX, furthermore, with appearance of the new FS, the dominant magnetic fluctuation changes from $(\pi,0)$ to $(\pi,\pi)$.
Thus, the self-energy shift by the correlation effect cannot be so large to reproduce experimental results.
This indicates that the Hund's coupling $J$ should not be so large.~\cite{rf:Nakamura}
The large outer FS observed in ARPES is probably composed of the $x^2-y^2$ orbital.
In addition, existence of the FS works in favor of superconductivity.
Thus, it will be natural that the outer FS composed of $x^2-y^2$ orbital exists above $z^2$ orbital around $\Gamma'$.
The self-energy shift in question works to equalize electron number included in each orbital.
Although this fact seems to be physically correct as the correlation effect, we have to investigate magnetic and superconducting properties without changing the character of the FSs from the band calculation.
This trouble is because the band calculation already includes partially exchange-correlation effect.
We should not overcount the static correlation effect at $\omega=0$ in the normal self-energy.
The final result which includes the correlation effect should reproduce the band character obtained by band calculations, which coincides with experimental results.
At present, we hereafter shift site energies of $z^2$ and $x^2-y^2$ orbitals by $+0.1$ and $+0.3$, respectively.~\cite{rf:Shift}
This tentative method allows us to investigate superconductivity without changing the character of FSs and the dominant magnetic fluctuation.
However, we need further investigation to proceed to more quantitative explanation.

\begin{figure}[ht]
\begin{center}
\includegraphics[height=40mm]{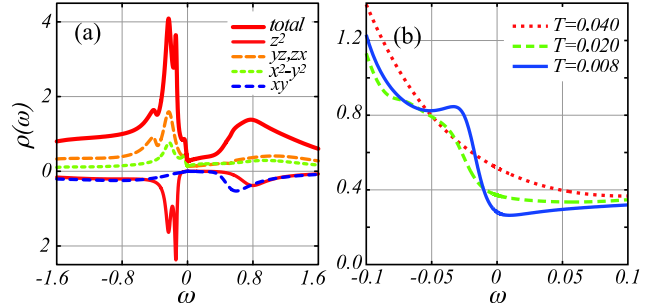}
\end{center}
\caption{(a) Partial DOS and (b) temperature dependence of total DOS at electron-doped $n=6.12$ for $(U,J,T)=(1.44,0.24,0.01)$.}
\end{figure}
Now, let us show numerical results in FLEX with site energies shifted.
In Fig.3, we display the partial DOS and its temperature dependence.
The $yz$, $zx$ and $x^2-y^2$ orbitals hold large weight at the Fermi level, and high DOS just below the Fermi level.
Corresponding to this fact, we find the pseudogap behavior in Fig.3(b).
On the contrary, we cannot find such behavior in the hole-doped region, although we do not demonstrate it.
These behaviors are consistent with the photoemission spectra,~\cite{rf:Sato} qualitatively.
Thus, this pseudogap can be naturally explained from the band structure.
This is the same reason as that proposed by Yada et al. for Na$_{0.35}$CoO$_2$.~\cite{rf:Yada}

Next, let us evaluate the NMR relaxation rate $1/T_1$, given by
$1/{T_1T}\propto \mathrm{Im}\chi^s_{\rm loc}(\omega)/\omega\big|_{\omega \to 0}.$
Using the Pade approximation, we carry out numerically analytic continuation for the local spin susceptibility, 
$\chi^s_{\rm loc}=\sum_{\mib q}\sum_{\ell,m}\chi^s_{\ell\ell,mm}(q)$, obtained in FLEX.
\begin{figure}[ht]
\begin{center}
\includegraphics[height=42mm]{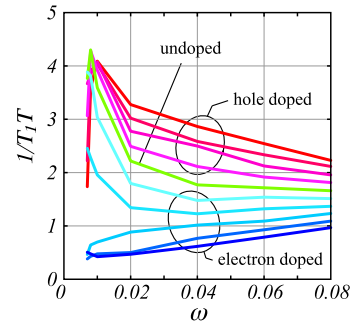}
\end{center}
\caption{NMR $1/T_1T$ at intervals of $\Delta n=0.04$ for $(U,J)=(1.44,0.24)$. The pseudogap appears in the electron-doped.}
\end{figure}
As illustrated in Fig.4, with decreasing temperatures, $1/T_1T$ is enhanced in the undoped and hole-doped systems.
On the other hands, in the electron-doped system, it shows the pseudogap behavior.
It becomes remarkable with electron-doping.
This qualitatively explains the experimental result in NMR.~\cite{rf:Grafe,rf:Ning,rf:Terasaki,rf:Nakai,rf:Mukuda,rf:Imai}
Such pseudogap behavior exists even without electron correlation in the present band structure.
The electron correlation only reduces the energy and temperature scale to actual values.

\begin{figure}[ht]
\begin{center}
\includegraphics[height=42mm]{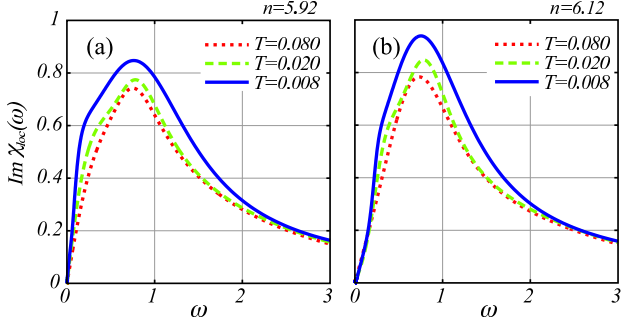}
\end{center}
\caption{$\mathrm{Im}\chi^s_{\rm loc}(\omega)$ at hole-doped $n=5.92$ (a) and electron-doped $n=6.12$ (b) for $(U,J)=(1.44,0.24)$.}
\end{figure}
In Fig.5, we show temperature dependence of imaginary part of the local spin susceptibility, $\mathrm{Im}\chi^s_{\rm loc}(\omega)$.
The slope at $\omega=0$ corresponds to $1/T_1T$.
We find that differences between the electron- and the hole-doped systems are restricted to the low frequency region, and the peak position at $\omega \sim 1$ is almost unchanged, although the spectral weight is broad in the hole-doped.
This fact may be related to weak sensitivity of $T_\mathrm{c}$ for carrier doping.

Next, we solve the superconducting Eliashberg equation for the spin-singlet pairing within FLEX.
The most probable candidate is the $s^\pm$-wave state.
The second largest eigenvalue is given by $d_{x^2-y^2}$-wave.
\begin{figure}[ht]
\begin{center}
\includegraphics[height=32mm]{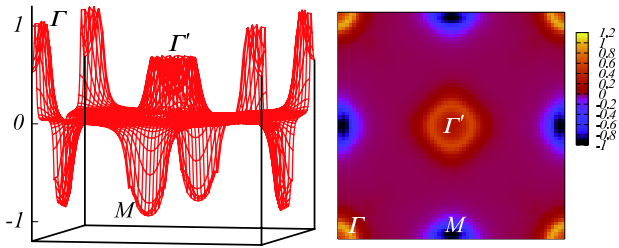}
\end{center}
\caption{Sum of the third- and fourth-band anomalous Green's functions at $n=6.00$ for $(U,J,T)=(1.44,0.24,0.008)$.\cite{rf:Del,rf:Del2}
It has about $+1$ on the outer FS around $\Gamma$, $-1$ around $M$ and $+0.5$ around $\Gamma'$.
The second-band possesses about $+1$ on the inner FS around $\Gamma$, and the first- and the fifth-band contributions are very small, although we does not display them.}
\end{figure}
In Fig.6, we display the sum of the third- and fourth-band anomalous Green's functions,~\cite{rf:Del} which indicates no nodal structure on the FSs more clearly than the gap function itself.
It has large weight on the outer FS around $\Gamma$ and the FSs around $\Gamma'$ and $M$.
The sign around $\Gamma$ ($\Gamma'$) is opposite to that around $M$.
In addition, we notice that magnitude of the gap function around $\Gamma'$ is about half.~\cite{rf:Del2}
This is consistent with magnitude of the gap on the outer FS obtained in the ARPES analysis.~\cite{rf:Ding,rf:Kondo}

\begin{figure}[ht]
\begin{center}
\includegraphics[height=42mm]{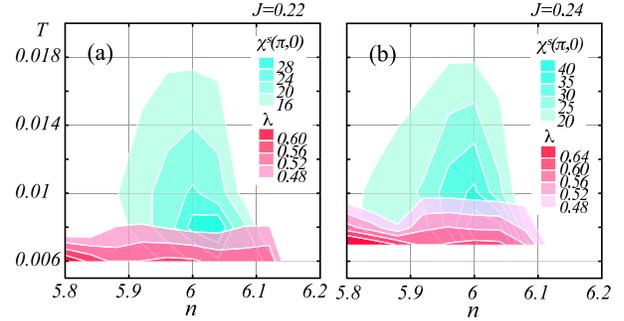}
\end{center}
\caption{Phase diagram for $J=0.22$ (a) and $0.24$ (b) at $U=1.44$.~\cite{rf:Phase}
$\lambda$ denotes the eigenvalue for the $s^\pm$-wave.}
\end{figure}
In Fig.7, we illustrate contour map of eigenvalues for the $s^\pm$-wave state and the static spin susceptibility $\chi^s({\mib Q},0)$ at ${\mib Q}=(\pi,0)$ for $J=0.22$ and $0.24$.
With carrier doping, the AF spin fluctuation decreases, while eigenvalues for $J=0.22$ are not so sensitive.
Eigenvalues for $J=0.24$ have stronger correlation with the AF spin fluctuation.~\cite{rf:Phase}
The phase diagram for small $J$ seems to be consistent with experiments,~\cite{rf:Kamihara,rf:Wen} although we need to evaluate $T_\mathrm{c}$ itself actually.
This point is related to the pairing mechanism discussed below.

Since the early stage, the pairing mechanism has probably been thought to be AF spin fluctuation,~\cite{rf:Mazin,rf:Kuroki} which is related to the AF state in the undoped system.~\cite{rf:Cruz}
Surely, the AF spin fluctuation is important.
However, this cannot explain insensitivity of $T_\mathrm{c}$ for carrier doping~\cite{rf:Kamihara,rf:Wen} and weak correlation with the AF fluctuation.~\cite{rf:Nakai,rf:Terasaki}
We need to make clear the pairing mechanism.

We hereafter study the mechanism within RPA, since the normal self-energy is unnecessary for this analysis.
First, let us examine spin and charge fluctuations, since the pairing interaction can be divided into the spin sector, $3\hat V^s\hat\chi^s\hat V^s/2$, and the charge sector, $\hat V^c\hat\chi^c\hat V^c/2$.
In addition, we can carry out analysis like perturbation theory without vertex corrections by truncating $\chi^{s,c}$ within finite-order terms.
In {Table~I}, we indicate the result for the $s^\pm$-wave in the linearized gap equation.
\begin{table}[ht]
\caption{Eigenvalues in the truncated perturbation at $n=6.1$ for $(U,J,T)=(1.00,0.10,0.008)$. $V^s$, $V^c$ and Total represent, respectively, eigenvalues with only spin sector, only charge sector, and the total contribution. Order $\infty$ corresponds to RPA.}
\begin{center}
\begin{tabular}{cccc}
\hline 
Order &   $V^s$   &   $V^c$  &   Total \\
\hline
    2   &   0.190   &   -0.430  &   0.050 \\
    3   &   0.375   &    1.520  &   1.790 \\
    4   &   0.525   &   -5.100  &  -4.900 \\
    5   &   0.640   &    15.80  &   15.80 \\
$\infty$&   0.953   & -0.0236  &   0.9379 \\
\hline
\end{tabular}
\end{center}
\end{table}
Within the third- and fifth-order perturbation, the spin sector and the charge sector work cooperatively, and the latter has dominant contribution.
However, in the second- and the fourth-order, the charge sector works negative.
Thus, the contribution of charge sector oscillates in each order term.
In such a case, by summing up to infinite order terms like RPA, we can expect to obtain the correct result.
Consequently, the charge sector does not contribute to the pairing interaction.
The dominant contribution comes from the spin sector.

Next, let us investigate individual matrix elements of $\hat\chi^s$ sandwiched between $\hat V^s$.
Even if components including $xy$ and $z^2$ orbitals are set to zero, the eigenvalue is almost unchanged.
On the other hand, the interaction among $yz$, $zx$, $x^2-y^2$ orbitals is especially important.
These three components are entangled completely.

\begin{figure}[ht]
\begin{center}
\includegraphics[height=44mm]{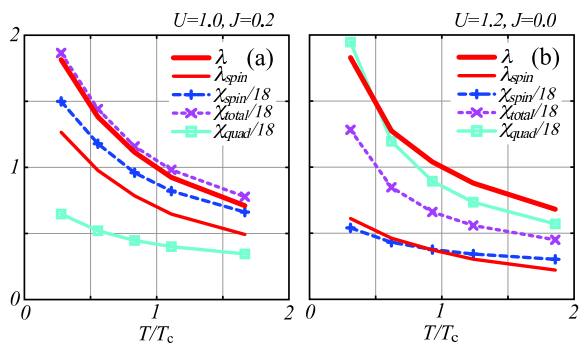}
\end{center}
\caption{Correlation between the eigenvalue $\lambda$ and some dominant fluctuations for (a) $(U,J)=(1.0,0.2)$ and (b) $(1.2,0.0)$ within RPA.~\cite{rf:Consistency}
$\lambda_{\rm spin}$ represents its weight which can be explained by only the AF spin fluctuation.
$\chi_{\rm spin}=\chi^s(\pi,9\pi/64)$, and $\chi_{\rm total}$ is the sum of $\chi_{\rm spin}$ and $\chi^{sQ}$.
$\chi_{\rm quad}$ represents $\chi^{QQ}$.~\cite{rf:Quadrupole}}
\end{figure}
In Fig.8, we show correlation between the eigenvalue $\lambda$ and dominant AF fluctuations for $J=U/5$ and $0$.
$\lambda_{\rm spin}$ represents the contribution from only spin-spin correlation $\chi_{\ell\ell,mm}$.
It can be obtained from the eigenvalue when any other components of $\hat\chi^s$ in the spin sector are set to zero.
For large $J=U/5$ in Fig.8(a), $60 \sim 70\%$ of eigenvalue can be explained by only spin-spin correlation, and the pairing interaction is dominated by the AF spin fluctuation. 
In this case, the behavior of $\lambda$ well correlates with the AF spin susceptibility $\chi^s({\mib q}\simeq{\mib Q},0)$.
This is inconsistent with experimental results.~\cite{rf:Kamihara,rf:Nakai,rf:Wen,rf:Terasaki}
On the contrary, for small $J=0$ in Fig.8(b), the spin-spin correlation leads to nothing but $\sim 30\%$ of eigenvalue, and the main part of the pairing interaction comes from the other fluctuation.
Correlation between $\lambda$ and the AF spin fluctuation is weak.
Instead, the fluctuation like $\chi^s_{\ell\ell,\ell m}+\chi^s_{\ell\ell,m\ell}$ becomes important.
This is the correlation $\chi^{sQ}$ between spin ($S_{\ell\ell}$) and spin-quadrupole ($Q_{\ell m}$).~\cite{rf:Quadrupole}
Independent of $J$, the behavior of $\lambda$ well correlates with $\chi_{\rm total}=\chi_{\rm spin}+\chi^{sQ}$.
Since $\lambda$ is a measure of the pairing interaction, this indicates that these fluctuations work cooperatively as the pairing interaction.
Furthermore, it is interesting that the spin-quadrupole fluctuation $\chi^{QQ}$ is enhanced for $J=0$, although what is more important as the pairing interaction is $\chi^{sQ}$.

In summary, we investigated the five-band Hubbard model for the iron-oxypnictide superconductors mainly by using FLEX.
We found the pseudogap behavior in the electron-doped region.
The most probable superconducting pairing state is the sign-changed $s$-wave.
It is proper that the Hund's coupling $J$ is not so large.
In this case, correlation between $T_\mathrm{c}$ and the AF spin fluctuation is weak, and then $T_\mathrm{c}$ will be insensitive to carrier doping, rather correlates with the AF spin-quadrupole fluctuation $\chi^{QQ}$.
The dominant contribution to the pairing interaction comes from correlation between spin and spin-quadrupole $\chi^{sQ}$ as the alternative to the AF spin fluctuation.
This novel pairing mechanism naturally comes from the fact that the system has many degrees of freedom in multiorbital system, and three bands which trigger unconventional superconductivity are strongly entangled.

\section*{Acknowledgement}
I thank Y. Matsuda, T. Shibauchi, K. Ishida and S. Fujimoto for stimulating my interest in this work.
I thank K. Yamada, T. Nomura, S. Onari and R. Arita for valuable discussions.
This work is supported by the Grant-in-Aid for the Scientific Research on Priority Areas (Grant No. 20029014) and the Global COE Program "The Next Generation of Physics, Spun from Universality and Emergence" from MEXT, Japan.



\begin{thebibliography}{99}
\bibitem{rf:Shift} These values themselves are not so important.
Qualitative results in this paper do not change, as far as we can reproduce the band structure near the FS.
\bibitem{rf:Del}
With the unitary matrix $u^{\mib k}_{\ell n}$, which diagonalizes the unperturbed Hamiltonian in the orbital representation,
we can obtain the band-diagonal $\mathcal{F}_{nn}(k)=\sum_{\ell m}u^{\mib k}_{\ell n}u^{-{\mib k}}_{mn}\mathcal{F}_{\ell m}(k)$.
Although we show the sum of the third- and fourth-band for want of space, they have large weight on the different FSs.
\bibitem{rf:Del2}
To examine the case with the large outer FS, we here show the result at $n=6.00$ not electron-doped region.
The magnitude of gap around $\Gamma'$ changes dependent on carrier doping.
It is $\sim 0.5$ in the hole-doped and $\sim 1$ in the electron-doped.
\bibitem{rf:Phase}
Increase of eigenvalues around hole-doped $n\simeq 5.80$ comes from change of the Fermi surface due to the self-energy shift.
It probably does not reflect the real fact in this system.
\bibitem{rf:Consistency}
In both cases, the $s^\pm$-wave state has the largest eigenvalue.
At a glance, this seems to be inconsistent with the result in Ref.\ref{rf:Yanagi}.
This is probably because of different dispersion relations, especially, size of the FS around $\Gamma'$.
\bibitem{rf:Quadrupole} 
For $yz$ ($2$) and $zx$ ($3$) orbitals, for instance, spin-quadrupole $Q^z_{23}$ is defined as $\sum_{{\mib k}\sigma} \sigma(c^\dagger_{{\mib k} 2\sigma}c_{{\mib k}+{\mib q} 3\sigma}+c^\dagger_{{\mib k} 3\sigma}c_{{\mib k}+{\mib q} 2\sigma})$, and then, $\chi^{sQ}_{23}=\langle\langle S^z(q),Q^z_{23}(-q) \rangle\rangle$, and $\chi^{QQ}_{23}=\langle\langle Q^z_{23}(q),Q^z_{23}(-q) \rangle\rangle$.
In Fig.8, for simplicity, we use $\chi^{sQ}=\chi^s_{22,23}+\chi^s_{22,32}+\chi^s_{23,22}+\chi^s_{32,22}$ at ${\mib q}=(\pi,9\pi/64)$, included in $\chi^{sQ}_{23}+\chi^{Qs}_{23}$, as the dominant fluctuation, and $\chi_{\rm quad}=\chi^{QQ}_{23}(\pi,9\pi/64)$.
For small $J$, $\chi^{sQ}_{\ell m}$ and $\chi^{QQ}_{\ell m}$ with $\ell,m \in$ $yz$, $zx$, or $x^2-y^2$ become large at ${\mib q}\simeq{\mib Q}$, and they show similar temperature dependence.
Although orbital fluctuations like $\chi^{s,c}_{\ell m,\ell m}-\chi^{s,c}_{\ell m,m \ell}$ are also enhanced in this case, these are more suppressed by introducing finite $J$ as compared with $\chi^{sQ}$ and $\chi^{QQ}$.
\end{thebibliography}
\end{document}